%% file: root.tex
\documentclass[letterpaper, 10 pt, conference]{ieeeconf}  

\IEEEoverridecommandlockouts                              

\usepackage{graphicx} 
\graphicspath{ {./figures/} }
\usepackage{amsmath} 
\usepackage{amssymb}
\usepackage{algorithm}
\usepackage[noend]{algpseudocode}
\usepackage{xcolor}
\definecolor{tumblue}{RGB}{0,101,189}

\usepackage{stfloats}
\usepackage{setspace}
\usepackage{tabularx}
\usepackage{booktabs}
\usepackage[T1]{fontenc}
\usepackage[scaled=0.85]{beramono}
\usepackage{listings}
\usepackage{caption}
\usepackage{cleveref}
\usepackage{chemformula}
\usepackage{url}
\usepackage{physics}
\Crefname{lstlisting}{Listing}{Listings}
\lstdefinestyle{sparqlstyle}{
  language=OCL,
  numbers=left,
  basicstyle=\footnotesize,
  stepnumber=1,
  numbersep=10pt,
  tabsize=2,
  showspaces=false,
  breaklines=true
}
\usepackage[nonumberlist,nopostdot,acronym,toc]{glossaries}
\usepackage{soul}
\usepackage{cases}
\usepackage{bm}
\usepackage{booktabs} 
\usepackage{siunitx}  

\usepackage{balance} 

\title{\LARGE \bf
Energy-Aware Integrated Proactive Maintenance \\ Planning and Production Scheduling
}

\author{Hongliang Li$^{1}$, Herschel C. Pangborn$^{2}$, and Ilya Kovalenko$^{3}$
\thanks{$^{1}$Hongliang Li is with the Department of Industrial and Manufacturing Engineering, The Pennsylvania State University, University Park, PA, USA, e-mail: hjl5377@psu.edu.}%
\thanks{$^{2}$Herschel C. Pangborn is with the Department of Mechanical Engineering, The Pennsylvania State University, University Park, PA, USA, e-mail: hcpangborn@psu.edu.}%
\thanks{$^{3}$Ilya Kovalenko is with the  Department of Mechanical Engineering and the Department of Industrial and Manufacturing Engineering, The Pennsylvania State University, University Park, PA, USA, e-mail: iqk5135@psu.edu.}%
}

\begin{document}

\maketitle
\thispagestyle{empty}
\pagestyle{empty}

\begin{abstract}

Demand-side energy management, such as the real-time pricing (RTP) program, offers manufacturers opportunities to reduce energy costs by shifting production to low-price hours. However, this strategy is challenging to implement when machine degradation is considered, as degraded machines have decreased processing capacity and increased energy consumption. Proactive maintenance (PM) can restore machine health but requires production downtime, creating a challenging trade-off: scheduling maintenance during low-price periods sacrifices energy savings opportunities, while deferring maintenance leads to capacity losses and higher energy consumption. To address this challenge, we propose a hierarchical bi-level control framework that jointly optimizes PM planning and runtime production scheduling, considering the machine degradation.
A higher-level optimization, with the lower-level model predictive control (MPC) embedded as a subproblem, determines PM plans that minimize total operational costs under day-ahead RTP. At runtime, the lower-level MPC executes closed-loop production scheduling to minimize energy costs under realized RTP, meeting delivery targets.
Simulation results from a lithium-ion battery pack assembly line case study demonstrate that the framework strategically shifts PM away from bottlenecks and high-price hours, meeting daily production targets while reducing energy costs.

\end{abstract}

\section{Introduction}
\label{sec:Intro}

\input{1_Intro}

\section{Problem Statement}
\label{sec:Problem}

\input{2_Problem_Statement}


\section{System Models}
\label{sec:Model}

\input{3_System_Model}


\section{Hierarchical Control Formulation}
\label{sec:HMPC}

\input{4_HMPC}

\section{Case Study}
\label{sec:Case}

\input{5_Case_Study}
\section{Conclusion}
\label{sec:Conclusion}

\input{6_Conclusion}
\bibliographystyle{IEEEtran}
%

\bibliography{bib_ACC}

\end{document}

%% file: 1_Intro.tex
The manufacturing sector accounts for roughly 30\% of global energy use~\cite{IEA2023}.
Since electricity cost represents about 40\% of primary production, manufacturers are highly motivated to curb their electricity spending~\cite{foster2020energy}.
Demand-side energy management (DSM) can reduce these costs by shifting energy-intensive activities to off-peak hours~\cite{10260642}.
One prominent DSM program is real-time pricing (RTP), in which electricity prices vary with grid conditions, including demand and renewable availability~\cite{elma2017implementation}.
Under RTP, manufacturers can proactively reschedule production to low-price periods, reducing utility charges without additional capital investment. 
In practice, however, RTP-based scheduling is complicated by machine degradation. 
Machine health evolves due to wear, fatigue, and operating stress~\cite{zhou2016energy}. 
Degradation undermines RTP benefits in two ways: (1) effective capacity declines, limiting flexibility to shift loads, and (2) energy consumption rises as health deteriorates, eroding savings even during low-price hours~\cite{luan2018trade}.
Therefore, operators face a trade-off between meeting deliveries at high energy cost and risking missed targets while waiting for off-peak prices.

\begin{figure}[t]
    \centering
    \includegraphics[width=1\linewidth]{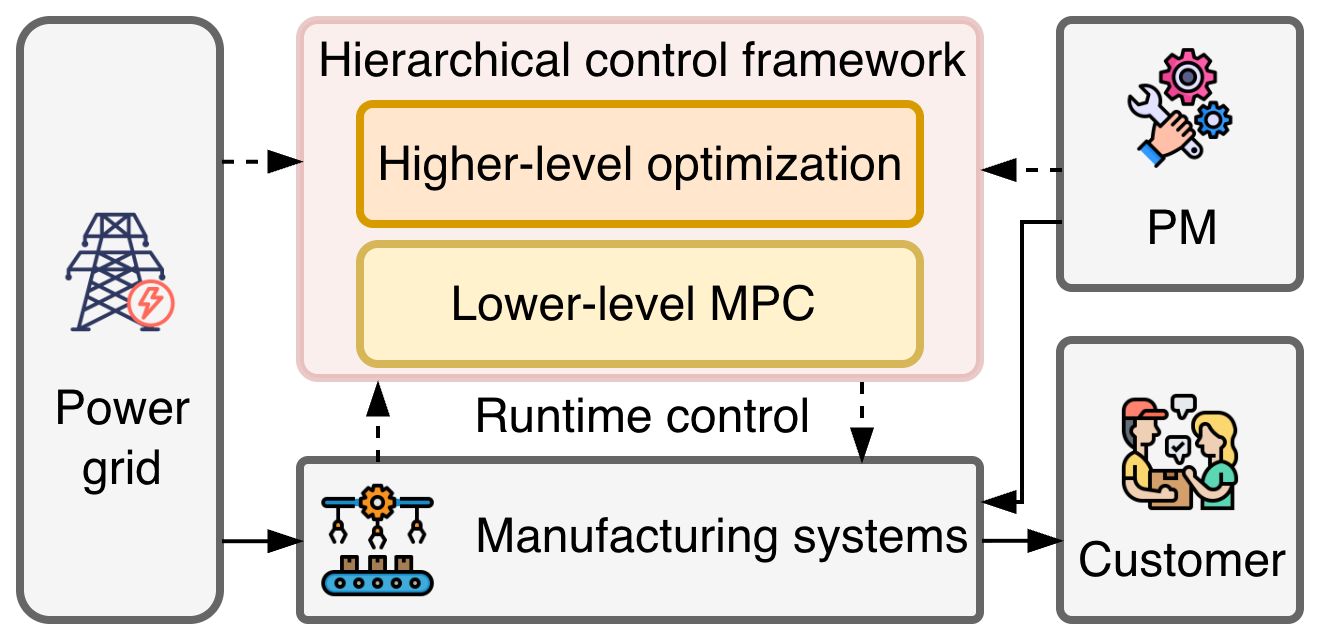}
    \caption{The proposed hierarchical control framework coordinates PM planning with runtime production scheduling, achieving system-level energy-efficiency improvements and energy cost savings. Dashed arrows denote information flow, while solid arrows denote maintenance execution, energy supply, and product delivery.}
    \label{fig:intro_framework}
\end{figure}

Proactive Maintenance (PM) restores machine health before
failures occur~\cite{kaihara2010proactive}.
However, PM requires scheduled downtime, interrupting production and reducing system flexibility.
Scheduling maintenance during low-price hours sacrifices opportunities for cost-effective production, whereas deferring maintenance risks operating with degraded capacity and reduced efficiency~\cite{susto2014adaptive}.
PM planning is further constrained by resource and time limits.
Balancing these factors yields a coupled planning problem that requires co-optimizing production scheduling and PM decisions with joint awareness of energy prices and machine health.
Existing literature has not fully addressed this intersection.
Most RTP scheduling formulations assume static health, treating machine capacity as time-invariant and decoupled from energy efficiency~\cite{li2025energy,li2025real}.
PM optimization typically prioritizes reliability, spare parts, and workforce costs without co-optimizing against energy efficiency~\cite{xia2022collaborative}. 
The few integrated approaches often adopt sequential decision-making and do not consider the bidirectional connections between production (driving degradation) and PM (altering future capacity and efficiency)~\cite{bencheikh2022approach}.
Another challenge is the inherent temporal mismatch, as PM is planned over long horizons, while production must react on short horizons to accommodate runtime plant operations.

Hierarchical control provides an effective framework for coordinating multi-timescale decisions by considering them at different control layers~\cite{scattolini2009architectures}.
A hierarchical model predictive control (MPC) has been applied to a renewable-powered ammonia plant, using a slow-timescale layer for storage and process control and a fast-timescale layer to manage the variability of the energy resource~\cite{tully2025hierarchical}.
However, implementing hierarchical control is challenging: effective coordination requires models that capture the connections among machine degradation, PM actions, and production dynamics, such as health-dependent processing capacity and energy consumption.

This paper proposes a hierarchical bi-level control framework that jointly optimizes PM planning and production scheduling under RTP, as shown in Fig.~\ref{fig:intro_framework}.
The higher level optimizes the PM plan offline by solving a bi-level optimization that accounts for day-ahead RTP with the lower-level MPC embedded as a subproblem to evaluate the production cost and feasibility of each candidate plan. 
At runtime, the lower level executes hourly production scheduling in closed loop under actual RTP, ensuring delivery targets and health-adjusted constraints.
The main contributions are:
(1) a unified state-space model that links buffer/machine dynamics with health evolution and energy use,
(2) a hierarchical bi-level control framework that coordinates the PM planning and production scheduling, and
(3) a case study that demonstrates the proposed control framework.
The remainder of the paper is organized as follows.
Section~\ref{sec:Problem} provides the assumptions and problem statement.
Section~\ref{sec:Model} describes the system model.
Section~\ref{sec:HMPC} details the hierarchical control framework.
Section~\ref{sec:Case} presents case study results.
Section~\ref{sec:Conclusion} provides conclusions and future work.

%% file: 2_Problem_Statement.tex
\begin{figure*}[t]
    \centering
    \includegraphics[width=0.95\linewidth]{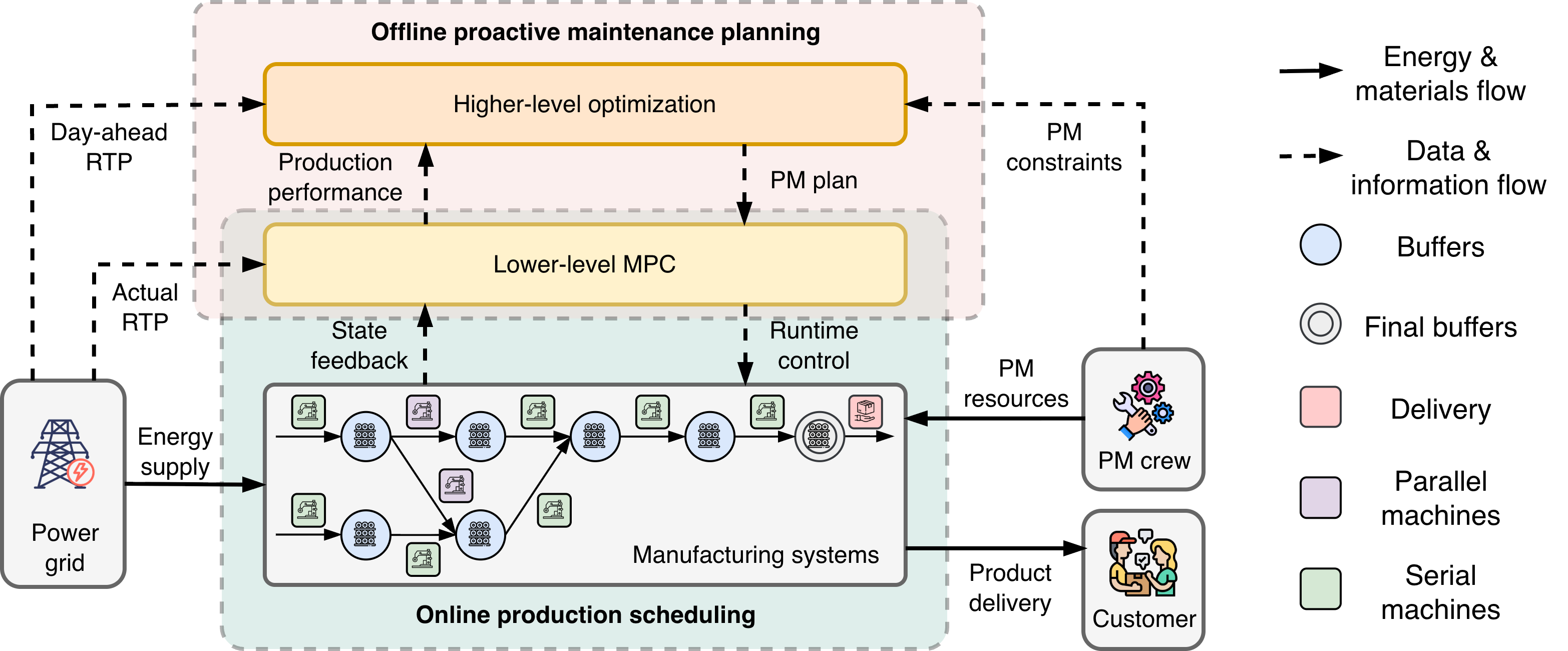}
    \caption{Hierarchical control framework for integrated PM planning and production scheduling with RTP-based energy costs in a make-to-order manufacturing environment.}
    \label{fig:framework}
\end{figure*}

In this study, we consider a multi-stage batch manufacturing system in which raw materials progress through a network of machines and buffers.
The system operates in a make-to-order (MTO) supply chain with known daily customer orders~\cite{zhai2022lead}.
Machines undergo progressive degradation: as health declines, effective processing capacity and energy efficiency decrease.
To preserve reliability, PM is employed. While maintenance restores health, it requires downtime that temporarily halts production.
Energy costs for processing are charged at the RTP provided by the utility company.
RTP programs generally include two profiles: a day-ahead forecast (RTP\textsubscript{est}), released 24 hours in advance for assisting end-users' energy management, and an actual real-time price (RTP\textsubscript{act}), updated hourly to determine utility charges~\cite{PJM_website}.
The offline PM planning uses the RTP\textsubscript{est} and the online production schedule responding to the RTP\textsubscript{act}.
The objective is to determine a PM plan and production schedules that (1) meet customer orders on time, (2) maintain machine reliability, and (3) minimize total energy costs.

To focus on the key decision aspects, we adopt the following modeling assumptions.
(1) Raw materials are continuously available. In MTO settings, procurement/preparation occurs based on confirmed orders.
(2) Compared with processing energy, the energy use directly associated with PM is negligible and thus omitted. PM incurs fixed costs reflecting maintenance labor and the opportunity cost of lost production.
(3) Energy consumption due to buffer holding is neglected, since machine processing dominates total energy use.
(4) As health deteriorates, we assume effective capacity decreases, and power draw increases linearly~\cite{schulze2022integration}.
We focus on gradual, predictable degradation and neglect sudden or random failures that may occur. Future work will relax this assumption to account for stochastic degradation.

%% file: 3_System_Model.tex
\subsection{Manufacturing Network Model}
Adapted from our prior work~\cite{li2025energy}, the manufacturing network is modeled as a discrete-time state space model:
\begin{equation}
\label{eq:ss_model}
    x(k+1) = A_s\,x(k) + B_s\,u(k) + W_s\,d(k)
\end{equation} where $x(k) \in \mathbb{R}^{n_x}$ denotes the buffer inventory vector, $u(k) \in \mathbb{R}^{n_u}$ represents the machine processing rates, and $d(k) \in \mathbb{R}^{n_p}$ captures the product outflow demands at time step $k$. The indices $n_x, n_u, n_p$ denote the total numbers of buffers, machines, and products, respectively. 
Buffers, machines, and products are indexed by $i$, $j$, and $p$, respectively.
The state-transition matrix $A_s \in \mathbb{R}^{n_x \times n_x}$ is taken as the identity matrix. The input matrix $B_s \in \mathbb{R}^{n_x \times n_u}$ encodes the manufacturing network topology:
\begin{equation}
\label{eq:b_ij}
    B_{s, ij} =
    \begin{cases}
        \phantom{-} 1, & \text{if $u_j$ delivers material into $x_i$} \\
        -1, & \text{if $u_j$ draws material from  $x_i$} \\
        \phantom{-} 0, & \text{otherwise}
    \end{cases}
\end{equation}
The matrix $W_s \in \mathbb{R}^{n_x \times n_p}$ links product outflows to buffer:
\begin{equation}
\label{eq:w_ip}
W_{s,ip} =
\begin{cases} 
-1, & \text{if $d_p$ is shipped from $x_i$} \\ 
\phantom{-} 0, & \text{otherwise} 
\end{cases}
\end{equation}
To track material withdrawals, we define the outflow matrix $B_o$ as $B_{o,ij} = -1$ if $B_{s,ij} < 0$, otherwise $B_{o,ij} = 0$.

\subsection{Machine Degradation Model}
\label{subsec:health}
\subsubsection{Machine Health Degradation}
Let $h(k)\in[0,1]^{n_u}$ denote the machine-health vector at time $k$,
and let $m(k)\in\{0,1\}^{n_u}$ denote the PM activation vector with $m_j(k)=1$ indicating maintenance on machine $j$ at time $k$.
Let $h_j(k)=1$ denote that the machine is in an “as-new” condition, and $h_j(k)=0$ denotes that the machine is in a failure condition.
A linear degradation model is adopted as a first-order approximation that captures usage-driven wear~\cite{rokhforoz2021distributed}:
\begin{equation}
\label{eq:health_evolution}
  h(k{+}1) = A_h \,h(k) - B_h \,u(k) + W_h\,m(k)
\end{equation}
where matrix $A_h \in \mathbb{R}^{n_u \times n_u}$ is taken as the identity matrix.
$B_h=\mathrm{diag}(b_1,\ldots,b_{n_u})$ and
$W_h=\mathrm{diag}(w_1,\ldots,w_{n_u})$ with $b_j,w_j\ge 0$.
Here, $b_j$ is the degradation rate per unit processing on machine $j$, representing the incremental health loss from each unit of production throughput.
The parameter $w_j$ is the health restoration rate per unit maintenance time on machine $j$, characterizing how maintenance activities recover machine condition.
Both parameters are machine-specific, reflecting differences in mechanical complexity, component wear characteristics, and maintainability.

\subsubsection{Maintenance Constraints}
PM is limited by crew availability and permissible working hours.
Over a planning horizon (e.g., one day), we have:
\begin{align}
    m_j(k) &\le \eta(k) \label{eq:window}\\
    \sum_{j=1}^{n_u} m_j(k) &\le N_{\text{crew}}(k)\label{eq:crew}
\end{align}
where $m_j(k)\in\{0,1\}$ indicates whether machine $j$ undergoes PM at time $k$,
$\eta(k)\in\{0,1\}$ encodes maintenance-permissible time windows, and
$N_{\text{crew}}(k)\in\mathbb{Z}_{\ge 0}$ denotes the available PM crew capacity
(e.g., the number of simultaneous maintenance tasks that can be staffed).
For example, if operators perform maintenance only during working hours
$T_{\mathrm{work}}=\{9,\dots,18\}$, then we have:
\begin{equation}
  \eta(k) =
  \begin{cases}
    1, & \text{if } k \in T_{\mathrm{work}}\\
    0, & \text{otherwise}
  \end{cases}
\end{equation}
PM requires human operators and induces production downtime, incurring both labor and opportunity costs.
We model the instantaneous maintenance cost as
\begin{equation}
    C_M(k) = \varphi \,\lVert m(k) \rVert_1
\end{equation}
where $\lVert m(k)\rVert_1$ counts the number of concurrent maintenance actions, and $\varphi>0$ is the per–machine–hour PM cost (labor, tooling, and lost throughput).

\subsubsection{Performance-Degradation Integration}
Machine degradation impacts both effective processing capacity and energy efficiency.
Let $\bar{\mathcal C}\in\mathbb{R}^{n_u}_{+}$ denote the as-new capacity vector and
$\kappa\in[0,1]^{n_u}$ the fractional capacity loss at zero health.
We model the health–capacity relation as:
\begin{equation}
\label{eq:capacity}
  \mathcal C(k)
   =
  \bar{\mathcal C}\,\odot\!\Big[\mathbf{1} - \kappa \odot \big(\mathbf{1}-h(k)\big)\Big]
\end{equation}
where $\odot$ denotes the elementwise product and $h(k)\in[0,1]^{n_u}$ is the health vector. Machine processing capacity is linearly decreased when the machine deteriorates.
Similarly, let $\bar{\mathcal E}\in\mathbb{R}^{n_u}_{+}$ denote the as-new power (or specific
energy-intensity) vector and $\gamma\in\mathbb{R}^{n_u}_{\ge 0}$ quantify the per-machine
energy penalty as health declines. We use a linear health–energy relation:
\begin{equation}
\label{eq:energy}
  \mathcal E(k)
   =
  \bar{\mathcal E}\,\odot\!\Big[\mathbf{1} + \gamma \odot \big(\mathbf{1}-h(k)\big)\Big]
\end{equation}

\subsection{Energy Consumption Model}
At each time step, the processing energy is:
\begin{equation}
\label{eq:energy_demand}
  E(k) = \Delta t\, \mathcal E(k)^\top u(k)
\end{equation}
where $\Delta t$ is the step length.
PM actions do not explicitly consume energy in the model, but they restore health and hence reduce future energy needs.
Let $\rho(k)$ denote the nominal RTP. Then the energy cost at time $k$ is:
\begin{equation}
\label{eq:energy_cost}
  C_E(k) = \rho(k)\,E(k)
\end{equation}
Note that for offline planning, the day-ahead RTP\textsubscript{est}, denoted as $\rho_\text{est}$, is used to estimate the energy cost.
For runtime control, the realized RTP\textsubscript{act}, denoted as $\rho_\text{act}$, is used for calculating the actual energy costs.

\subsection{System Operational Constraints}
\label{subsec:constraints}
\subsubsection{Machine and Buffer Capacity Constraints}
Buffer levels $x(k)\in\mathbb{R}^{n_x}$ are bounded by physical limits:
\begin{equation}
\label{eq:x_capacity}
  x_{\min} \le x(k) \le x_{\max}
\end{equation}
Machines are limited by health-dependent capacities and are unavailable during PM.
Then, the control input satisfies:
\begin{equation}
\label{eq:u_capacity}
  \mathbf{0} \le u(k) \le \mathcal C(k)\,\odot\big(\mathbf{1}-m(k)\big)
\end{equation}
so that when $m_j(k)=1$, we have $u_j(k)=0$.

\subsubsection{Production Flow Conservation}
Material balance requires sufficient inventory for processing and deliveries.
A conservative (no shortage) condition is:
\begin{equation}
\label{eq:flow}
  x(k) + B_o\,u(k) + W_s\,d(k) \ge \mathbf{0}
\end{equation}
Cumulative shipments $\pi(k)$ over the prediction horizon $N$ must track the production target $\sigma(k)$:
\begin{equation}
\label{eq:production}
  \pi(k) + \xi(k) \ge \sigma(k), \qquad \xi(k) \ge 0
\end{equation}
where $\xi(k)$ is the slack variable that allows temporary, penalized deviations to preserve feasibility. Cumulative shipments $\pi(k)$ can be calculated as:
\begin{equation}
  \pi(k) = \sum_{t=0}^{k} d(t), \qquad k=0,\dots,N{-}1 
\end{equation}

%% file: 4_HMPC.tex
Fig.~\ref{fig:framework} shows the hierarchical control architecture. 
The higher level solves an offline PM planning problem, formulated as a bi-level optimization, whose inner problem is the lower-level MPC (L-MPC) that evaluates the production optimality and feasibility of a candidate PM plan.
The lower-level runs a closed-loop receding-horizon MPC that uses real system feedback and actual RTP to generate runtime production schedules and track production targets.
The hierarchical structure operates across different timescales such that hourly production scheduling in the lower level respects the daily PM plan from the higher level.

\subsection{Lower-level MPC}
Given a PM plan $m$ and a production target $\sigma$, the L-MPC seeks a production schedule that balances inventory regulation, energy cost, smooth operation, and target tracking.
Before linearization, the energy term renders the problem bilinear.
We state the nominal problem and then introduce a
convex McCormick relaxation.
Given a PM plan $m$, the nominal L-MPC is formulated as:
\begin{subequations}
\label{eq:LL-mpc}
\begin{align}
\min_{x,u,d,\xi}\quad
& J_L  = \sum_{k=0}^{N-1}\!\Big(
\|x(k)-x_g\|_{Q}^2
+ \lambda_e\, C_E(k) \notag \\
&+ \lambda_u\,\|u(k)-u(k{-}1)\|_2^2
+ \lambda_\xi\,\|\xi(k)\|_1
\Big) \label{eq:LL-obj} \\
\text{s.t.}\quad
& k \in \{0,1,\dots,N-1\} \label{eq:timeindex}\\
& x(0) = x_0, \; h(0) = h_0 \label{eq:init}\\
& x(k{+}1) = A_s x(k) + B_s u(k) + W_s d(k) \label{eq:dyn}\\
& h(k{+}1) = A_h h(k) - B_h u(k) + W_h m(k) \label{eq:ll-h}\\
& x_{\min} \le x(k) \le x_{\max} \label{eq:inv-bounds}\\
& 0 \le u(k) \le \mathcal{C}(k)\,\odot\big(1-m(k)\big) \label{eq:cap-LL}\\
& x(k) + B_o\,u(k) + W_s\,d(k) \ge 0 \label{eq:noshortage}\\
& d(k) \ge 0 \label{eq:ship}\\
& \pi(k) + \xi(k) \ge \sigma(k), \quad \xi(k)\!\ge\! 0 \label{eq:target-LL}
\end{align}
\end{subequations}
where $x_0$ and $h_0$ are the initial conditions, and $x_g$ is the safe inventory level, a design parameter that balances work-in-process reduction against production flexibility by penalizing excess buffer accumulation through the quadratic cost $\|x(k)-x_g\|_{Q}^2$.
The weights
$\lambda_e,\lambda_u,\lambda_\xi\!\ge\!0$ tune the trade-offs.
To model the integration of machine degradation and machine power while keeping the L-MPC convex, we use McCormick envelopes to relax the bilinear term of $C_E(k)$~\cite{mccormick_computability_1976}.
We introduce the auxiliary variable $w(k) := h(k)\odot u(k)$.
Then we have:
\[
E(k) = \big((\bar{\mathcal E}{+}\gamma)^{\!\top} u(k) - \gamma^{\!\top} w(k)\big)\Delta t
\]
The bilinearities appear through $w(k) = h(k)\odot u(k)$.
Given the machine health bound $0\!\le\!h_j(k)\!\le\!1$ and the machine capacity bound $0\!\le\!u_j(k)\!\le\!\bar{\mathcal C}_j$,
the convex McCormick envelope for $w_j(k)$ is:
\begin{align}
\label{eq:LL-mcc}
& w_j(k) \ge 0, 
  && w_j(k) \le u_j(k), \\
& w_j(k) \le \bar{\mathcal C}_j\,h_j(k), 
  && w_j(k) \ge u_j(k) - \bar{\mathcal C}_j\big(1 - h_j(k)\big) \notag
\end{align}
With \eqref{eq:LL-mcc}, the energy cost term becomes linear:
\begin{equation}
\label{eq:LL_energy_cost_lin}
C_E(k) = \rho(k)\,\Big((\bar{\mathcal E}{+}\gamma)^{\!\top} u(k) - \gamma^{\!\top} w(k)\Big)\Delta t
\end{equation}
and the L-MPC \eqref{eq:LL-mpc} reduces to a convex quadratic program.
Empirically, the McCormick envelope's relaxation is tight~\cite{boland2017bounding}. We also apply the bound $0\!\le\!u_j(k)\!\le\!\bar{\mathcal C}_j$ to ensure the machine capacity is within the as-new condition capacity.

\subsection{Higher-level Bi-level Optimization}
The higher level is formulated as a bi-level optimization problem with the L-MPC as the embedded subproblem to evaluate the system performance based on the candidate PM plan.
The bi-level mixed-integer linear problem is:
\begin{subequations}
\label{eq:HL-op}
\begin{align}
\min_{m,\theta}\quad
& J_H = \theta + \sum_{t=0}^{T-1}\!\Big(
-\lambda_p\,\rho_\text{est}\,\|m(t)\|_{1} + \lambda_m\,C_M(t)
\Big) \label{eq:HL_obj} \\
\text{s.t.}\quad
& t\in\{0,\dots,T-1\} \label{eq:HL_timeindex}\\
& m_j(t) \le \eta(t) \\ 
& \sum_{j=1}^{n_u} m_j(t) \le N_{\mathrm{crew}}(t) \qquad  \label{eq:HL_crew}\\
& 1 \ge h(T) \ge\; h^{\min} \label{eq:HL_health}\\
& \theta \ge\; J_L^*(m) \label{eq:HL_epi}
\end{align}
\end{subequations}
where $\eta(t)\!\in\!\{0,1\}$
enforces permissible windows, and $N_{\mathrm{crew}}(t)$ limits crew capacity.
The weight $\lambda_p$ tunes the preference of PM allocation during high RTP periods.
Weight $\lambda_m$ tunes the PM costs, primarily due to the crew cost and loss of production.
Constraint \eqref{eq:HL_health} ensures that the end-of-the-day machine health $h$ is higher then the safe machine health $h^{min}$ to prevent machine failure.
Constraint \eqref{eq:HL_epi} uses the optimal value $J_L^*(m)$ of the L-MPC
\eqref{eq:LL-mpc} evaluated under the plan $m$.

\begin{algorithm}[t]
\caption{GBD for PM Planning}
\label{alg:GBD-PM}
\begin{algorithmic}[1]
\Require Initial $(x_0,h_0)$, RTP $\rho(t)$, targets $\tau(t)$, weights $(\lambda_p,\lambda_m,\lambda_e,\lambda_u,\lambda_\xi)$, tolerance $\varepsilon$
\State $\mathcal{L}\gets\emptyset$ \Comment{cut set}
\State $\mathrm{UB}\gets+\infty$, $\mathrm{LB}\gets-\infty$, $\ell\gets0$
\Repeat
  \State \textbf{Master:} Solve \eqref{eq:HL-op} with cuts $\mathcal{L}$ to get $(m^{[\ell]},\theta^{[\ell]})$
  \State $\mathrm{LB}^{[\ell]}\gets$ master problem objective value
  \State \textbf{Subproblem:} Solve \eqref{eq:LL-mpc} at $m^{[\ell]}$ to obtain $J_L^*(m^{[\ell]})$ \\ \quad \quad \quad \quad \quad \quad \quad \, and duals $(\mu,\lambda^h)$
  \State \textbf{Cut:} Build $g(m^{[\ell]})$ via \eqref{eq:subgrad-LBB} and add \eqref{eq:cut-LBB} to $\mathcal{L}$
  \State \textbf{Update UB:} $\mathrm{UB}^{[\ell]}\gets$ min($\mathrm{UB}^{[\ell-1]}$, ~\eqref{eq:UB_update}) 
  \State $\ell\gets \ell+1$
\Until{$\mathrm{UB}^{[\ell-1]} - \mathrm{LB}^{[\ell-1]} \le \varepsilon$}
\State \Return $m^\star$, $\theta^\star$, $\mathrm{UB}$ 
\end{algorithmic}
\end{algorithm}

\subsection{Solution Approach}
The high-level problem~\eqref{eq:HL-op} evaluates a PM plan
$m \in \{0,1\}^{n_u \times T}$ via the L-MPC~\eqref{eq:LL-mpc}.
The L-MPC optimal value $J_L^*(m)$ enters the higher-level objective through a variable $\theta$ with the constraint $\theta \ge J_L^*(m)$.
As detailed in Alg.~\ref{alg:GBD-PM}, we adopt a Generalized Benders Decomposition (GBD) approach for solving the bi-level MILP~\eqref{eq:HL-op}~\cite{bolusani2022framework}.
GBD alternates a fixed-size master problem with repeatedly solving the convex subproblem. 
We denote the problem~\eqref{eq:HL-op} as the master problem with the embedded subproblem of L-MPC.
At iteration $\ell$, the master proposes $(m^{[\ell]},\theta^{[\ell]})$; solving L-MPC at $m^{[\ell]}$ yields the value $J_L^*(m^{[\ell]})$ and dual multipliers.
Let $\mu_j(k)\!\ge\!0$ denote the dual multiplier of the per–time-step capacity
constraint~\eqref{eq:cap-LL}
and let $\lambda^h(k)\in\mathbb{R}^{n_u}$ be the multiplier of the health dynamics~\eqref{eq:ll-h}
where $\tau(k)\in\{1,\dots,T\}$ maps the intra-day index $k$ to the day-ahead slot $t=\tau(k)$.

In the L-MPC~\eqref{eq:LL-mpc}, the PM plan $m$ affects the problem through~\eqref{eq:cap-LL} and~\eqref{eq:ll-h}.
Under convexity and strong duality of the L-MPC QP, the optimal value function $J_L^*(m)$ is convex in $m$, and a valid subgradient at $m$ is obtained from the dual solution as:
\begin{equation}
\label{eq:subgrad-LBB}
\begin{aligned}
g_j(k;\hat m) &= \mathcal C_j(k) \mu_j(k;\hat m)
- \big[W_h^\top \lambda^h(k;\hat m)\big]_j\\
\qquad
j &= 1,\dots,n_u,\;\quad k=0,\dots,N{-}1
\end{aligned}
\end{equation}
The associated global supporting hyperplane in the lifted
$(m,\theta)$ space is:
\begin{equation}
\label{eq:cut-LBB}
\theta
\;\ge\;
J_L^*(\hat m)\;+\;
\sum_{k=0}^{N-1}\sum_{j=1}^{n_u}
g_j\!\bigl(k;\hat m\bigr)\,
\bigl(m_j(\tau(k)) - \hat m_j(\tau(k))\bigr)
\end{equation}
Because the L-MPC problem is a convex
QP, strong duality holds and
\eqref{eq:cut-LBB} is valid for all $m\in\{0,1\}^{n_u\times T}$.
The GBD alternates a fixed-size higher-level MILP (master) with repeated solves of the convex L-MPC (subproblem).
At iteration $\ell$, the master solves~\eqref{eq:HL-op} subject to all accumulated cuts $\mathcal{L}$, producing a candidate PM plan $m^{[\ell]}$, an epigraph value $\theta^{[\ell]}$, and an updated lower bound $\mathrm{LB}^{[\ell]}$.
The subproblem L-MPC is then solved at $m^{[\ell]}$ to obtain the operational cost $J_L^*(\tilde m^{[\ell]})$ and the dual multipliers
$(\mu,\lambda^h)$.
The subgradient \eqref{eq:subgrad-LBB} and the cut \eqref{eq:cut-LBB} are formed and appended to $\mathcal{L}$.
The incumbent upper bound $\mathrm{UB}^{[\ell]}$ is updated by adding the subproblem
cost to cost function of~\eqref{eq:HL_obj} as:
\begin{equation}
\label{eq:UB_update}
    J_L^*(\tilde m^{[\ell]})
    + \sum_{t=0}^{T-1}\!\Bigl(
      -\lambda_p\,\rho(t)\,\|\tilde m^{[\ell]}(t)\|_1
      + \lambda_m\,\|\tilde m^{[\ell]}(t)\|_1
    \Bigr)
\end{equation}
The iteration terminates when:
\begin{equation}
\mathrm{UB}^{[\ell]}\;-\;\mathrm{LB}^{[\ell]}
\;\le\;
\varepsilon
\end{equation}
where $\varepsilon > 0$ is a prescribed relative convergence tolerance.

\subsection{Integration of the Hierarchical Control}
The proposed bi-level optimization-based hierarchical control enables bi-directional communication between higher-level PM planning and lower-level production scheduling.
The higher-level optimization passes the PM plan to the lower-level MPC, which enforces maintenance-induced downtime through the capacity constraint~\eqref{eq:cap-LL} and receives health restoration through the health dynamics~\eqref{eq:ll-h}.
The PM plan $m$ thus shapes the lower-level feasible region by dictating when machines are available and at what health states they operate.
The lower-level MPC provides feedback to the higher level through the bi-level optimization structure.
It should be noted that, in the online implementation, the lower-level MPC executes in closed loop with hourly state feedback, but does not modify the daily PM plan once it is committed.
This reflects practical scheduling realities: maintenance crews, spare parts, and tooling are arranged at the start of the day based on the PM plan, making intra-day rescheduling operationally costly.
The GBD-based solution approach ensures that the PM plans are evaluated against production performance through the embedded L-MPC.
Hence, the PM plan is production-aware even though it is not updated at runtime.
Future work could introduce a rolling-horizon re-planning mechanism that triggers higher-level PM plan updates when realized conditions deviate significantly from day-ahead assumptions, enabling tighter bi-directional coordination.

%% file: 5_Case_Study.tex
\begin{figure}[t]
    \centering
    \includegraphics[width=1\linewidth]{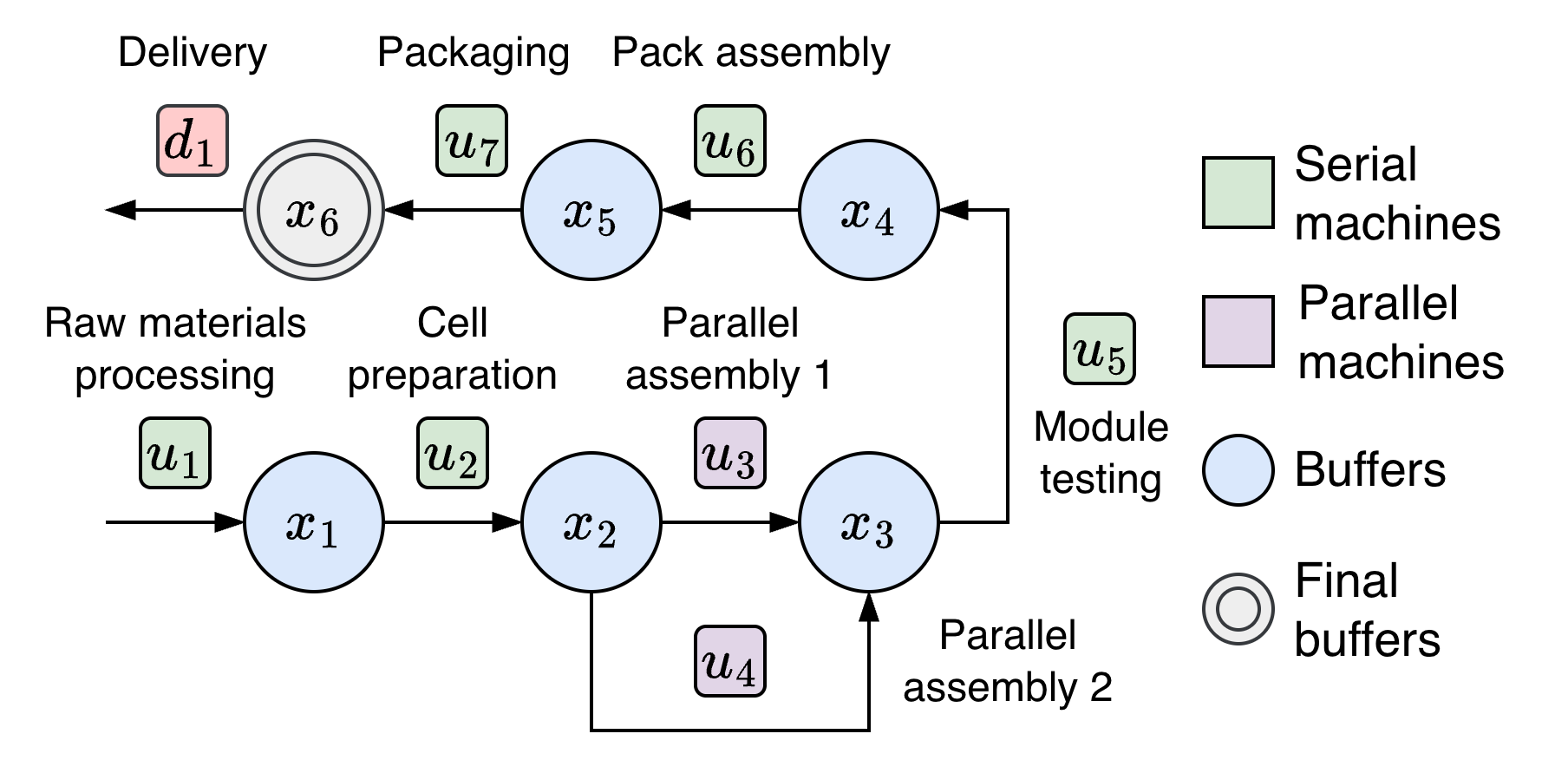}
    \caption{Lithium-ion battery pack manufacturing system
considered in the case study.}
    \label{fig:case_system}
\end{figure}

\begin{figure*}[t]
    \centering
    \includegraphics[width=1\linewidth]{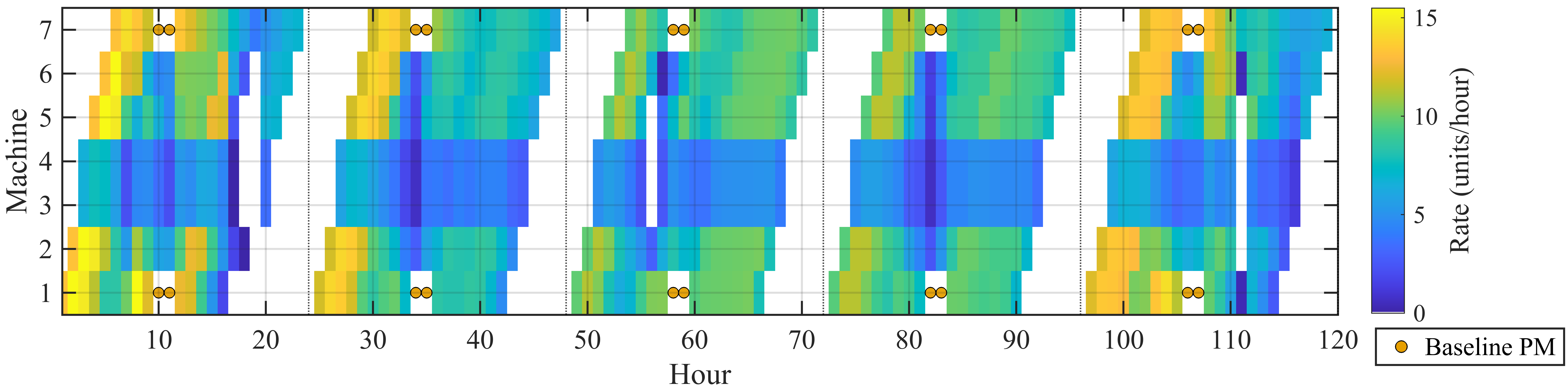}
    \caption{Machine schedules under the baseline controller. An orange dot indicates that the machine is under PM. }
    \label{fig:base_schedule}
\end{figure*}

\begin{figure*}[t]
    \centering
    \includegraphics[width=1\linewidth]{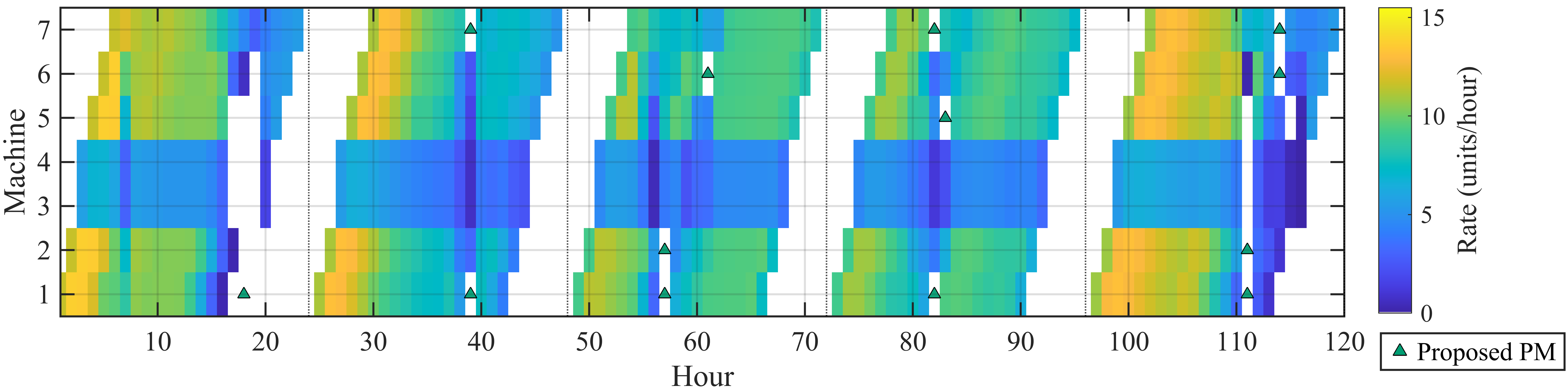}
    \caption{Machine schedules under the proposed framework. A green triangle indicates that the machine is under PM.}
    \label{fig:hmpc_schedule}
\end{figure*}

\begin{figure*}[t]
    \centering    \includegraphics[width=0.98\linewidth]{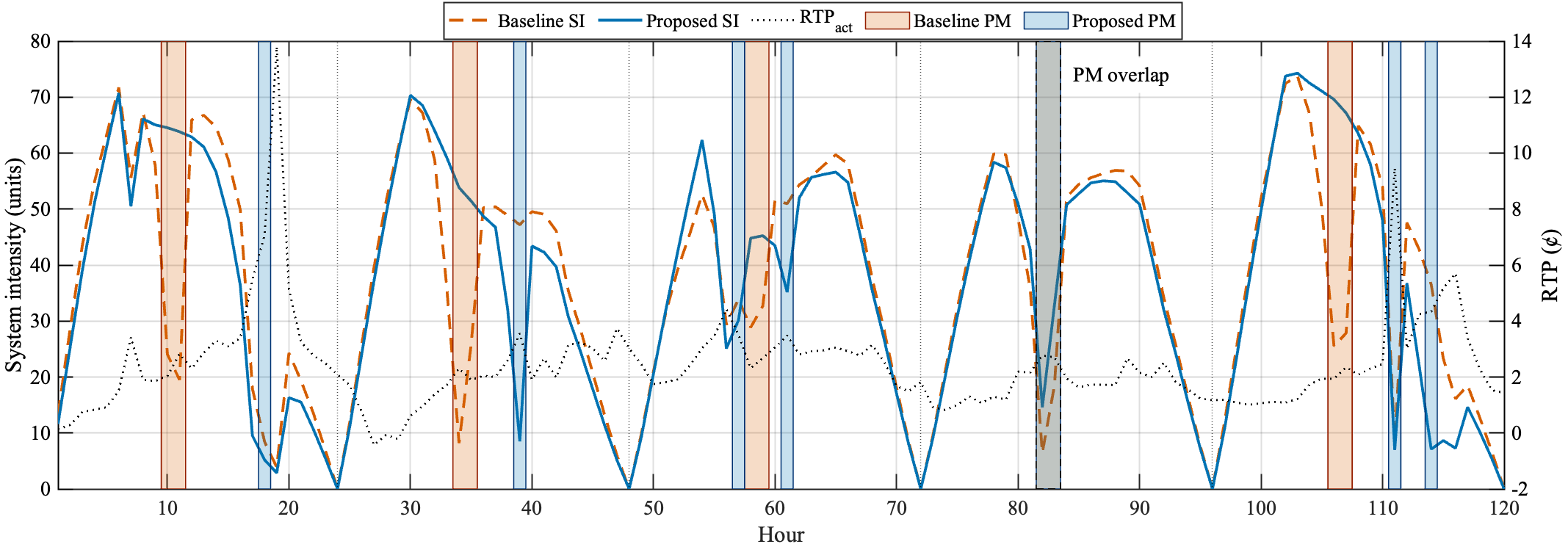}
    \caption{System intensity and PM plan of baseline controller and proposed framework.}
    \label{fig:intensity}
\end{figure*}

\subsection{Case Study Setup}
We evaluate the proposed hierarchical control framework in simulation on a Li-ion battery pack assembly line.
The production network consists of six buffer stages and seven processing machines arranged in a hybrid serial-parallel configuration, as illustrated in Fig.~\ref{fig:case_system}.
The production line follows a multi-stage assembly process beginning with raw material processing, progressing through cell preparation, and branching into parallel module assembly operations. The workflow continues through module testing, pack assembly, and concludes with final testing and packaging. This configuration represents typical battery manufacturing workflows~\cite{heimes2018lithium}.

The system configuration and parameters are based on~\cite{heimes2018lithium}.
$\Delta t$ is defined as 1 hour.
Buffer capacities are $x_{\max}=[300,250,200,180,180,400]^\top$ units, reflecting physical space.
We assume full capacity loss at zero health, that is, $\kappa =1$.
Machine capacities are
$\bar{\mathcal{C}}=[40,25,20,20,20,25,30]^\top$ units/hour, with $u_3$ and $u_4$ operating in parallel. 
Baseline (as-new) per-unit energy use is
$\bar{\mathcal{E}}=[0.5,0.9,1.8,1.8,2.6,1.0,0.4]^\top$ kWh/unit, and sensitivity parameters
$\boldsymbol{\gamma}=[0.1,0.3,0.8,0.8,1.0,0.4,0.2]^\top$ capture how energy intensity increases with load/health.
Inventory holding costs follow a quadratic penalty with $Q=\mathrm{diag}(0.04,0.06,0.08,0.10,0.10,1.00)$ with the increasing downstream to encourage lean work in process.
Safe inventory $x_g=0$ to minimize the work in process.
The L-MPC weight parameters are $\lambda_e = 1.0$, $ \lambda_u = 1.5$, and larger penality on the production fulfillment $ \lambda_\xi = 1\times10^{5}$.
Maintenance costs are $\varphi = \$ 100$/hour.
Machine health evolves with production-induced wear $B_h = 5\times10^{-4} I$, where $I$ is the identity matrix.
Maintenance restores health at machine-specific rates
$W_h = \mathrm{diag}(0.15,0.15,0.20,0.20,0.20,0.15,0.15)$ per hour; higher rates for $u_3$–$u_5$ reflect their bottleneck criticality.
The PM weighting parameters are $\lambda_p = 1.5$ and $\lambda_m = 10$.
Maintenance is workforce-constrained to 9:00–18:00 with at most two concurrent activities, and each machine must satisfy a day-end health floor of 60\% to ensure next-day operability.
Daily demand is 150 battery packs, consistent with just-in-time automotive requirements.
We compare the proposed framework to a baseline controller in which a fixed daily PM plan is passed to the L-MPC.
Based on the system configurations, the fixed PM plan is set as the daily one-hour PM for machines 1 and 7.
This represents common industrial practice where maintenance follows a fixed calendar-based policy.
Simulations use PJM Chicago-area RTP from May 2024 over a 5-day (120-hour) horizon~\cite{PJM_website}.

\begin{figure}[t]
    \centering
    \includegraphics[width=1\linewidth]{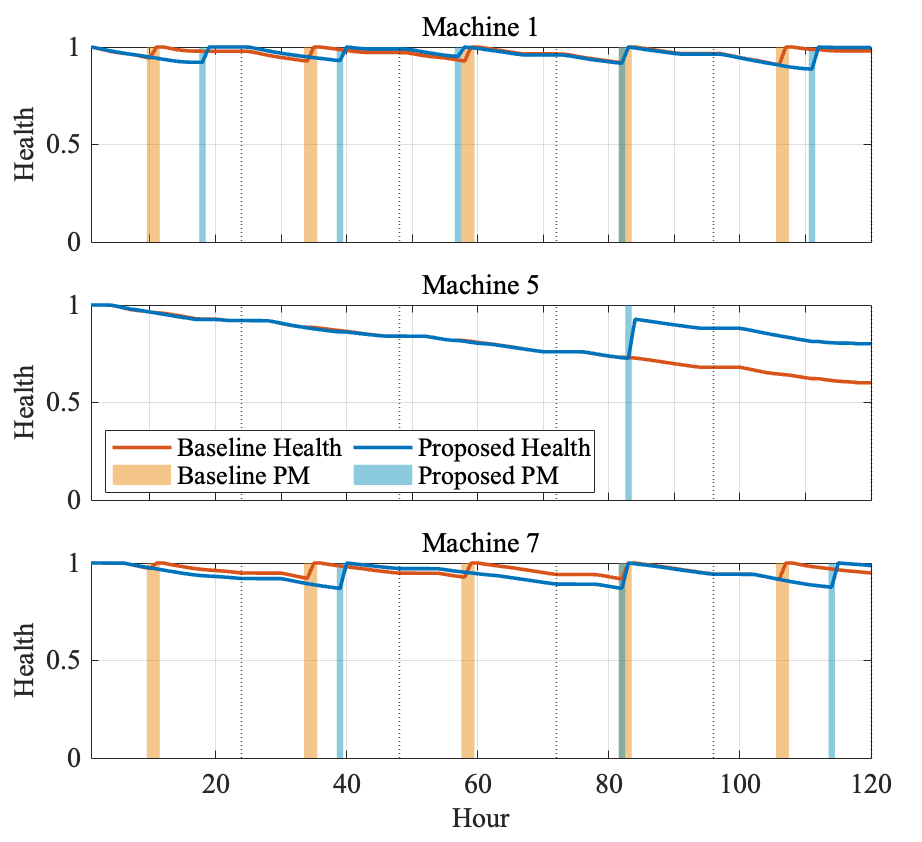}
    \caption{Machine health degradation and restoration under baseline controller and proposed framework.}
    \label{fig:health_states}
\end{figure}

\begin{figure}[t]
    \centering
    \includegraphics[width=1\linewidth]{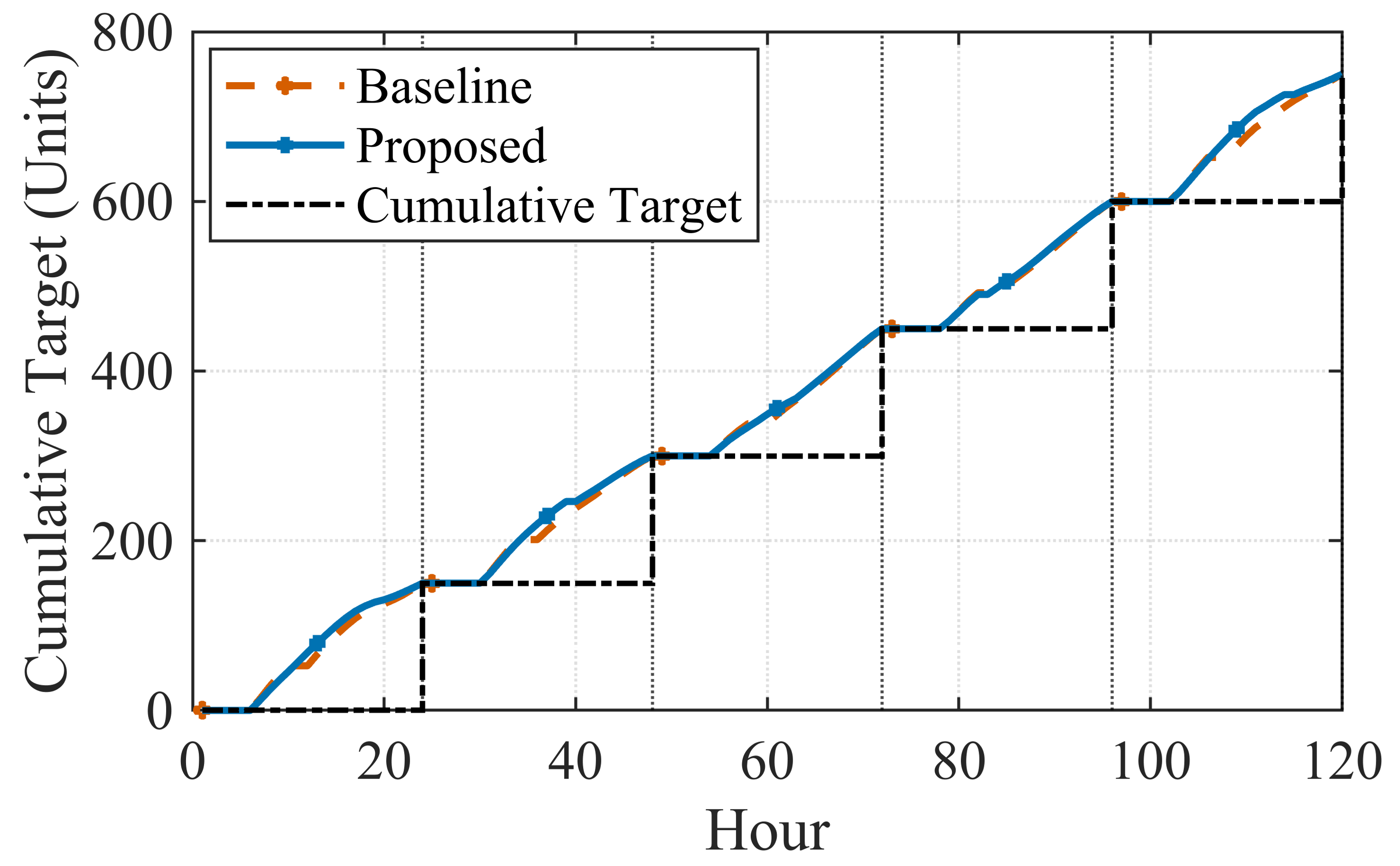}
    \caption{Product delivery and production goal under baseline controller and proposed framework.}
    \label{fig:goal}
\end{figure}

\subsection{Results and Discussions}
Fig.~\ref{fig:base_schedule} and Fig.~\ref{fig:hmpc_schedule} compare machine-runtime schedules under the baseline controller with the fixed PM plan and the proposed hierarchical control framework.
In the baseline controller, a fixed PM plan is applied independently of the production schedule: machines 1 and 7 each receive one hour of daily maintenance at a predetermined time within the working window, regardless of RTP conditions, buffer states, or machine health levels.
The fixed PM interrupts production periodically. Consequently, the L-MPC must both avoid high-RTP hours and work around PM downtime, forcing higher processing rates in the remaining windows.
For example, at hour 20, when RTP is elevated, machines under the baseline controller still accelerate to meet the daily production goal, sacrificing energy-saving opportunities.
The proposed framework integrates PM planning with production scheduling via the embedded L-MPC, explicitly accounting for production performance.
The resulting PM is preferentially scheduled near higher-RTP periods, preserving low-RTP hours for production and thereby increasing energy savings.
The proposed framework also reflects the coupling between degradation and workload. PM is planned after intensive production to restore health and mitigate machine failure risk.
Fig.~\ref{fig:intensity} presents system intensity, PM, and RTP. System intensity (SI) $I_s (k) $ is defined as the sum of machine processing rates at time $k$, namely $I_s (k) = \Delta t \sum_{j=1}^{n_u} u_j(k)$.
For both the baseline controller and the proposed framework, the L-MPC responds to RTP to avoid peak prices.
However, the baseline exhibits sharper swings in system intensity due to uncoordinated PM interruptions.
The proposed hierarchical control framework yields a smoother production profile by aligning PM with high-RTP intervals, thereby improving energy efficiency and benefiting long-term machine health.

Fig.~\ref{fig:health_states} presents the selected machine health degradation profiles under both the baseline controller and the proposed hierarchical control framework.
The baseline approach follows a rigid maintenance policy, resulting in abrupt health restorations at fixed intervals, regardless of actual machine condition or production requirements. 
This leads to premature maintenance in some cases and delayed intervention in others.
Fig.~\ref{fig:goal} shows actual daily production versus the production goal. Both controllers meet the cumulative shipment targets each day. 
The proposed framework, however, tracks more tightly during target transitions and reaches intermediate checkpoints earlier, reflecting faster post-PM recovery and better use of low-RTP periods.
Table~\ref{tab:kpi-compare-extended} summarizes economic performance over the 120-hour horizon.
The proposed framework reduces total operational cost by 23.70\% relative to the baseline, driven primarily by optimized maintenance scheduling that cuts preventive maintenance cost by 35\% and total maintenance time from 20 to 13 hours. Despite reaching production targets, energy cost decreases by 5.38\%, and cost per unit drops by 5.41\%.
The proposed framework achieves a 3.45\% improvement in average system health, demonstrating its ability to simultaneously lower operating costs and enhance machine reliability. 

\begin{table}[t]
\centering
\setlength{\tabcolsep}{4pt}
\caption{Performance evaluation}
\label{tab:kpi-compare-extended}
\begin{tabular*}{\columnwidth}{@{\extracolsep{\fill}}lrrr@{}}
\toprule
\textbf{Metric} & \textbf{Baseline} & \textbf{Proposed} & \textbf{Diff (\%)}\\
\midrule
Energy Cost (\$)      & 1232.91   & 1166.62   & -5.38    \\
PM Cost (\$)         & 2000.00   & 1300.00   & -35.00   \\
Total Cost (\$)       &3232.91    & 2466.62   & -23.70   \\
Cost/Unit (\$/Unit)   & 16.44     & 15.55     & -5.41    \\
Avg. Health           & 0.87      & 0.90      & +3.45    \\
PM (Hours)           & 20        & 13        & -35.00   \\
\bottomrule
\end{tabular*}
\end{table}

\subsection{Scalability Analysis}
All simulations were performed on an Apple MacBook Pro with an M3 Max processor and 48 GB of RAM, using MATLAB~\cite{MATLAB2024} with YALMIP~\cite{lofberg2004yalmip} and Gurobi~\cite{gurobi}.
The 5-day (120-hour) simulation completed in 36.32~s of total wall-clock time.
The L-MPC had a mean solve time of 0.25~s per hourly decision and a worst-case solve time of 0.47~s. This is well within the 1-hour control interval, confirming that the proposed hierarchical framework is computationally tractable for online deployment.
The day-ahead solution of the higher-level optimization took approximately 1.22~s, which  is negligible relative to its 24-hour planning cycle.

Regarding scalability to larger systems, the L-MPC remains convex under the McCormick relaxation.
Each Benders cut is generated efficiently. 
The L-MPC scales with the number of machines, buffers, and horizon length and can be solved efficiently with commercial solvers that exploit sparsity and warm-starting.
The master MILP, however, contains $n_u \times 24$ binary variables for hourly PM decisions.
In the worst case, solving the master problem is NP-hard and scales exponentially with the binary dimension.
For the case study with 168 binary variables, the master problem solves within seconds, but for significantly larger systems (e.g., hundreds of machines or multi-line configurations), the master problem may become the computational bottleneck.
Potential mitigation strategies include aggregating PM decisions to coarser time blocks (e.g., per-shift rather than per-hour), exploiting warm-starting across GBD iterations, and applying decomposition techniques at the lower level to maintain tractability~\cite{wei2023scalable}.

%% file: 6_Conclusion.tex
This paper presented a framework for integrating PM scheduling with production planning in manufacturing systems, addressing the central challenge of coordinating equipment health management with DSM-driven operations. We developed a hierarchical control architecture with a two-level structure that explicitly embeds machine health dynamics into production decisions, moving beyond approaches that treat maintenance and production as separate problems. A Li-ion battery pack case study demonstrated the framework’s practical value, where it is able to meet production targets while reducing energy costs. By coupling health-dependent capacity constraints with time-varying electricity prices, the framework balances immediate production needs against long-term equipment sustainability.
Future work will model stochastic degradation using chance-constrained or robust optimization to improve reliability and will broaden the objective to include additional sustainability metrics, such as carbon emissions and water usage.